\documentclass[prc,aps,amsmath,amssymb,nofootinbib,superscriptaddress,12pt]{revtex4-2}
\usepackage{epsfig}
\usepackage[bookmarksnumbered,bookmarksopen,colorlinks,citecolor=blue,linkcolor=blue]{hyperref}
\usepackage{amsmath}
\usepackage{amssymb}
\usepackage{tipa}
\usepackage{float}

\begin{document}

    \title{ Universal empirical coupled channel calculations for capture cross sections in synthesis of superheavy nuclei }
 
\author{H. Yao}
\affiliation{ Department of Physics, Guangxi Normal University, Guilin 541004, China }

\author{Y. J. Duan}
\affiliation{ Department of Physics, Guangxi Normal University, Guilin 541004, China }

\author{N. Wang}
\email{wangning@gxnu.edu.cn}
\affiliation{ Department of Physics, Guangxi Normal University, Guilin 541004, China }
\affiliation{ Guangxi Key Laboratory of Nuclear Physics and Technology, Guilin 541004, 	China }
 
\author{D. D. Zhang}
\affiliation{ CAS Key Laboratory of Frontiers in Theoretical Physics, Institute of Theoretical Physics, Chinese Academy of Sciences, Beijing 100190, China } 

\author{X. T. He}
\affiliation{College of Material Science and Technology, Nanjing University of Aeronautics and Astronautics, Nanjing 210016, China } 

\author{T. L. Zhao}
\affiliation{Department of Physics, Hunan Normal University, Changsha 410081, China } 

\author{P. H. Chen}
\affiliation{School of Physical Science and Technology, Yangzhou University, Yangzhou 225009, China } 

\author{P. W. Wen}
\affiliation{China Institute of Atomic Energy, 102413 Beijing, China } 

\author{L. Zhu}
\affiliation{Sino-French Institute of Nuclear Engineering and Technology, Sun Yat-sen University, Zhuhai 519082, China } 

\author{X. J. Bao}
\affiliation{Department of Physics, Hunan Normal University, Changsha 410081, China } 

\author{J. J. Li}
\affiliation{College of Material Science and Technology, Nanjing University of Aeronautics and Astronautics, Nanjing 210016, China } 

\author{L. Guo}
\affiliation{School of Nuclear Science and Technology, University of Chinese Academy of Sciences, Beijing 100049, China }

     \begin{abstract}
        A universal empirical coupled channel model, ECC2, is proposed for the systematic description of capture cross sections in heavy-ion fusion reactions. The model combines a modified Broglia--Winther potential, an effective barrier curvature, and a deformation-dependent barrier distribution width. The most probable barrier heights predicted by ECC2 are in good agreement with the extracted values based on measured excitation functions for 367 systems, and the root-mean-square (rms) deviation is 1.39 MeV. ECC2 is applied to 25 fusion reactions ranging from $^{16}$O+$^{197}$Au to $^{54}$Cr+$^{248}$Cm. The predicted capture excitation functions are consistent with the EBD2.2 predictions for all systems and agree well with the available experimental data.
     \end{abstract}

     \maketitle

\newpage
  
\section{Introduction}


The synthesis of superheavy nuclei is fundamental to understanding the limits of nuclear existence and remains one of the foremost challenges in low-energy nuclear physics~\cite{Zhou_2020,Adamian_2020,Ogan15,Hof00}. Theoretical predictions of an ``island of stability" near the neutron magic number $N=184$ \cite{meldner1966predictions,Sobiczewski1966} have motivated extensive experimental programs, which have led to the synthesis of elements up to $Z=118$ through heavy-ion fusion reactions~\cite{Thoennessen2013}. These reactions proceed through three sequential stages: the capture of the projectile by the target, the formation of a compound nucleus, and its de-excitation through neutron evaporation and $\gamma$ emission. The capture cross section, which characterizes the probability that the colliding nuclei overcome the Coulomb barrier and become trapped in the nuclear potential pocket, is the essential first step for predicting the evaporation-residue cross section. The dinuclear system (DNS) model is widely used for this purpose in theoretical studies~\cite{Zagrebaev2001Synthesis,Adam04,Wang_2017,Rachkov2014,Li_2018,Li_2023,Wu_2025,Zhang_2018,Li_20181}.

\begin{figure}[htbp]
  	\setlength{\abovecaptionskip}{ 0 cm}
  	\includegraphics[angle=0,width=0.85\textwidth]{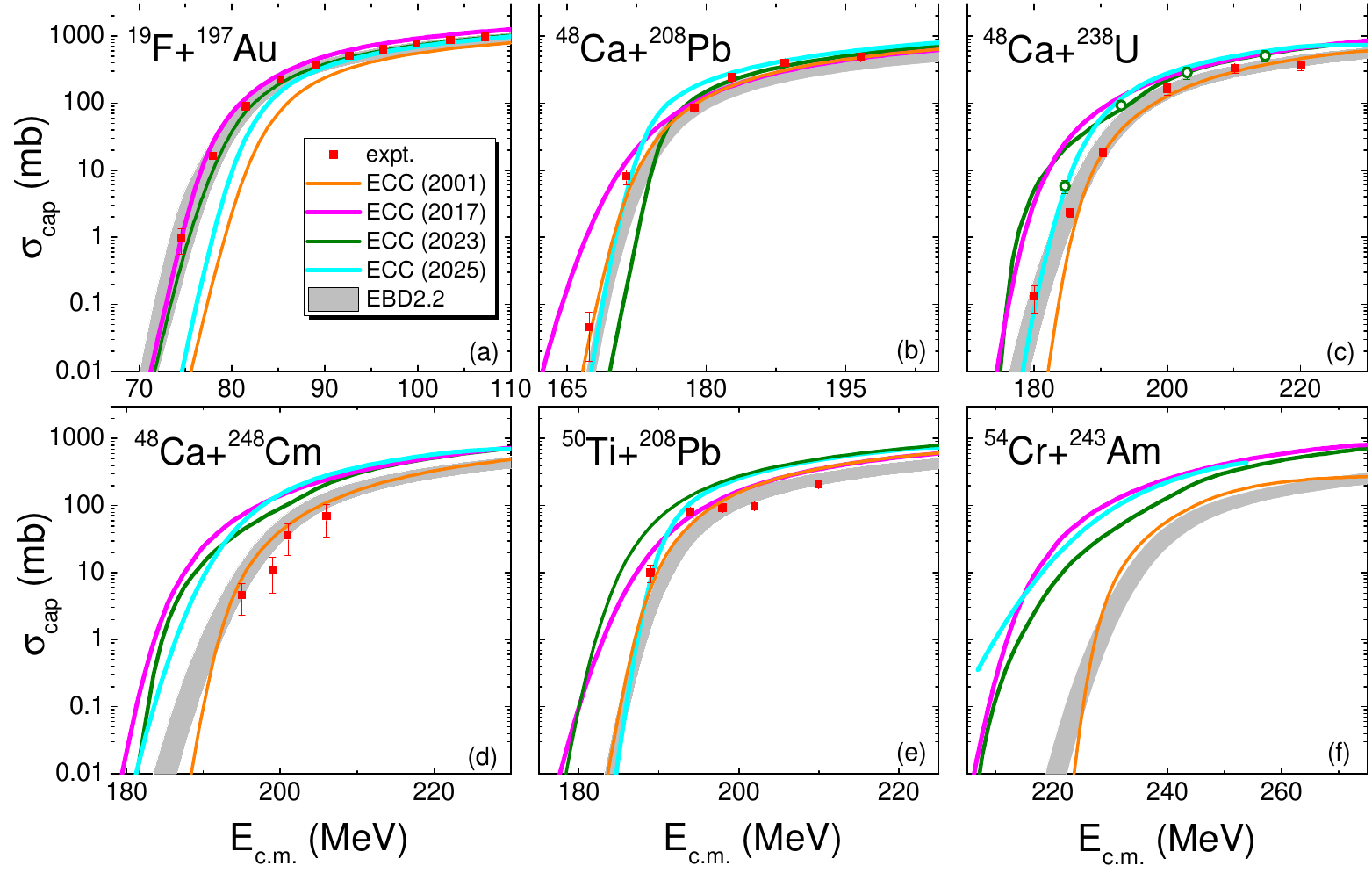}
  	\caption{Predicted capture excitation functions for six reactions: $^{19}$F + $^{197}$Au, $^{48}$Ca + $^{208}$Pb, $^{48}$Ca + $^{238}$U, $^{48}$Ca + $^{248}$Cm, $^{50}$Ti + $^{208}$Pb, and $^{54}$Cr + $^{243}$Am. The squares and circles denote the experimental data from Refs.~\cite{Hinde2002Role,Prokhorova_2008,Kozulin_2010,Itkis_2015,Clerc_1984,Itkis2022Experimental}, the shaded bands indicate the EBD2.2 predictions, and the curves denote the results obtained with different versions of the ECC model \cite{Zagrebaev2001Synthesis,Wang_2017,Li_2023,Wu_2025}.}
  	\label{fig:eccbenchmark}
  \end{figure}

  In the DNS model, the capture cross section is often calculated using the empirical coupled channel (ECC) method. In this method, the coupling of the projectile--target relative motion to the intrinsic degrees of freedom of the nuclei is described by an asymmetric Gaussian barrier distribution. Three key parameters characterize this distribution: the most probable barrier height $B_m$, the left width $\Delta_1$, and the right width $\Delta_2$. However, because different implementations employ different interaction potentials (such as the Woods--Saxon potential~\cite{Wang_2017} and the double-folding potential \cite{Li_2018,Li_2023,Wu_2025,Zhang_2018,Li_20181}), the parameters of different ECC versions vary significantly, leading to large uncertainties in the predicted capture cross sections for superheavy systems. As shown in Fig.~\ref{fig:eccbenchmark}, four different ECC versions \cite{Zagrebaev2001Synthesis,Wang_2017,Li_2023,Wu_2025} give markedly different results. For medium-mass systems, the different versions are broadly consistent with each other above the barrier, but diverge dramatically at sub-barrier energies, with cross sections differing by orders of magnitude. For heavy systems, the predictions differ drastically both above and below the barrier. The EBD2.2 predictions (shaded bands) are included as a reference. It is therefore necessary to develop a universal ECC model that provides a satisfactory description for fusion systems ranging from light to heavy.
 
  A further issue concerns the penetration coefficient. In traditional ECC models, it is calculated using the Hill--Wheeler formula based on a parabolic approximation of the barrier top. However, this approximation yields an excessively narrow barrier width because it neglects the long-range tail of the Coulomb potential. As a result, the sub-barrier penetration probability is systematically overestimated. An effective barrier curvature is thus required to describe the barrier width more realistically.
  
  This paper presents ECC2, a universal empirical coupled channel model designed to address these limitations. The model introduces three key improvements. First, the bare barrier is calculated from a modified Broglia--Winther potential, denoted BW91$^*$, in which a repulsive term is introduced to improve the capture-pocket depth while leaving the barrier height nearly unchanged. Second, the standard Hill--Wheeler curvature is replaced by an effective curvature that reproduces the broader width of a realistic Coulomb barrier and thereby corrects the overestimation of sub-barrier penetration. Third, the asymmetric Gaussian barrier distribution adopts the deformation-dependent width from the EBD2 model \cite{Wang_2025,Wang_2026}.

The article is organized as follows. Section~II formulates the ECC2 model, covering the BW91$^*$ potential, the effective barrier curvature, and the deformation-dependent barrier distribution. Section~III presents capture excitation functions for 25 fusion reactions ranging from $^{16}$O- to $^{54}$Cr-induced systems, in which the ECC2 predictions are compared with those of EBD2.2, the universal Wong formula (Fusion-v2), and ECC (2017). The corresponding barrier parameters are tabulated, and the global accuracy of the predicted barrier heights is assessed against experimental values. Section~IV gives a brief summary and outlook.

 \section{Empirical Coupled Channel Model}
 
 In this section, we firstly introduce the general formalism for capture cross section calculations. Then, the modified BW91$^*$ nuclear potential is presented in sub-section II.B, followed by the deformation-dependent asymmetric Gaussian distribution in sub-section II.C. Finally, the effective barrier curvature will be introduced in sub-section II.D.
 
 \subsection{Theoretical framework}

 The capture cross section at a given center-of-mass energy \( E_{\text{c.m.}} \) is expressed as the partial-wave sum
 \begin{equation}
 	\sigma_{\text{cap}}(E_{\text{c.m.}}) = \frac{\pi \hbar^2}{2\mu E_{\text{c.m.}}} \sum_J (2J + 1)T(E_{\text{c.m.}}, J),
 	\label{eq:capture}
 \end{equation}
  where \( \sigma_{\text{cap}} \) denotes the capture cross section, \( E_{\text{c.m.}} \) and \( J \) are the incident center-of-mass energy and the relative angular momentum, and \( T(E_{\text{c.m.}}, J) \) is the probability that the colliding nuclei overcome the Coulomb barrier in the entrance channel. The penetration probability depends on the nuclear potential, the barrier curvature, and the barrier distribution, each of which is discussed below.

To compute the capture cross section, the barrier penetration probability is required. In the single-barrier penetration model, the interaction potential around \( B_0 \) can be approximated by an inverted parabola, for which the analytical penetration probability is given by the well-known Hill--Wheeler formula~\cite{Wong73,Bass74}:
\begin{equation}
	T^{\text{HW}}(E_{\text{c.m.}}, J, B_0) = \frac{1}{1 + \exp\left\{-\frac{2\pi}{\hbar \omega}\left[E_{\text{c.m.}} - B_0 - \frac{\hbar^2}{2\mu R_B^2}J(J+1)\right]\right\}}.
	\label{eq:HW}
\end{equation}
In this expression, \(\hbar \omega \) denotes the curvature of the Coulomb barrier at sub-barrier energies and \(\mu\) denotes the reduced mass, $\mu = A_1 A_2/(A_1 + A_2)\,u$, with $A_1$ and $A_2$ the mass numbers of the projectile and target. $R_B$ denotes the average barrier radius.

To account for the coupling between the relative motion of the reaction partners and dynamical processes such as dynamical deformation and nucleon transfer, the total penetration probability in Eq.~\eqref{eq:capture} is averaged over the barrier height in the ECC calculations,
\begin{equation}
	T(E_{\text{c.m.}}, J) = \int f(B)T^{\text{HW}}[E_{\text{c.m.}}, J, B]\,dB,
	\label{eq:avg}
\end{equation}
where the barrier distribution function \(f(B)\) is approximately described by an asymmetric Gaussian form,
\begin{equation}
	f(B) = 
	\begin{cases} 
		\dfrac{1}{\mathcal{N}} \exp\left[-\left(\dfrac{B - B_m}{\Delta_1}\right)^2\right], & B \leqslant B_m, \\[10pt]
		\dfrac{1}{\mathcal{N}} \exp\left[-\left(\dfrac{B - B_m}{\Delta_2}\right)^2\right], & B > B_m.
	\end{cases}
\end{equation}
Here, \( f(B) \) satisfies the normalization condition $\int f(B)\, dB = 1$, with $\mathcal{N} = \sqrt{\pi} (\Delta_1 + \Delta_2)/2$ being the normalization coefficient.

 \subsection{Modified Broglia-Winther Potential}

  The nuclear potential adopted in this work is based on the Woods--Saxon parametrization proposed by Broglia and Winther (BW91), which is derived from the densities of the colliding nuclei and an effective two-body force~\cite{BW91,Reisdorf_1994}. In the present formulation, the potential barrier is calculated with the BW91 potential supplemented by a repulsive term $V_1$, 
 \begin{equation}
 	V_N(r) = \frac{V_0 + V_1}{1 + \exp\left(\frac{r - R_{N}}{a}\right)}.
 	\label{eq:VN}
 \end{equation}
Here,
 \begin{equation}
 	V_0 = -16\pi \frac{R_1 R_2}{R_1 + R_2} \gamma a,
 	\label{eq:V0}
 \end{equation}
 with \( a = 0.63 \) fm and $\gamma = 0.95 \left( 1 - 1.8 I_1 I_2 \right)$ MeV\,fm$^{-2}$, where $I_i=(N_i-Z_i)/A_i$ is the isospin asymmetry of the reaction partners ($i=1$ for projectile and $i=2$ for target). The radius parameter is defined as
 \begin{equation}
 	R_N = R_1 + R_2 + 0.29\ \text{fm},
 	\label{eq:R0}
 \end{equation}
 with 
 \begin{equation}
 	R_i = 1.233 A_i^{1/3} - 0.98 A_i^{-1/3} \text{ fm}.
 	\label{eq:Ri}
 \end{equation}
 
    \begin{figure}[htbp]
 	\setlength{\abovecaptionskip}{0.2cm}
 	\includegraphics[angle=0,width=0.6\textwidth]{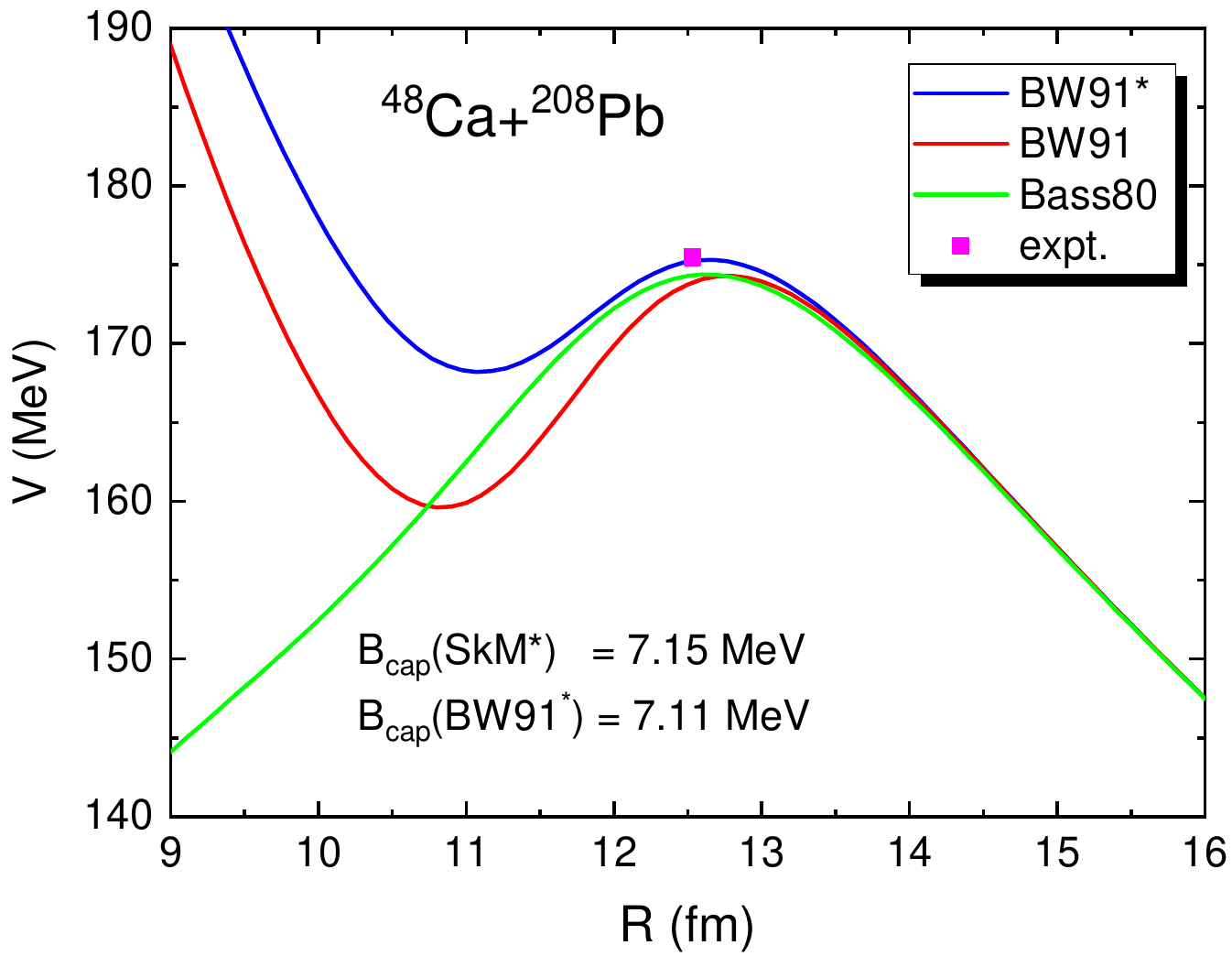}
 	\caption{Nucleus-nucleus potential $V$ for the reaction $^{48}$Ca + $^{208}$Pb. The red, blue, and green solid lines denote the original BW91 potential, the modified BW91$^*$ potential, and the Bass80 potential, respectively. The pink square represents the extracted barrier parameters~\cite{Chen23AtData}. The capture-pocket depths $B_{\rm cap}$ obtained from the SkM$^*$ energy-density functional and from the BW91$^*$ potential are also indicated.}
 	\label{fig:VRcompare}
 \end{figure}

  To better describe the Pauli-blocking effect at small internuclear distances, a repulsive term $V_1$ is introduced  to the original BW91 potential,
 \begin{equation}
 	V_1=1.34 \left(\frac{R_1 R_2}{R_1 + R_2}\right)^2 \left(\frac{R_N}{R}\right)^4 \text{ MeV}.
 	\label{eq:V1}
 \end{equation}
As a result of the repulsive term $V_1$, the barrier height $B_0$ remains almost unchanged, while the depth of the capture pocket $B_{\rm cap}$ in the modified Broglia--Winther (${\rm BW91^*}$) potential $V = V_N + V_C$ is brought into close agreement with that obtained from the Skyrme energy-density functional for the entrance-channel potential \cite{liumin}. The Coulomb potential is given by
\begin{equation}
	V_C = 
	\begin{cases} 
		\frac{Z_1 Z_2 e^2}{2(R_1+R_2)} \left( 3 - \frac{R^2}{(R_1+R_2)^2} \right) & \text{if } R \leq R_1+R_2, \\ 
		\frac{Z_1 Z_2 e^2}{R} & \text{if } R > R_1+R_2.
	\end{cases}
\end{equation}

Fig.~\ref{fig:VRcompare} presents the nucleus-nucleus potential for $^{48}$Ca + $^{208}$Pb obtained from BW91$^*$, BW91, and Bass80 \cite{Bass80}. In the outer barrier region, all three potentials are in good agreement; notably, BW91$^*$ gives the barrier height closest to the extracted data (pink square)~\cite{Chen23AtData}. At shorter distances, however, the three models diverge substantially: BW91 produces a pronounced pocket, whereas Bass80 exhibits no pocket at all. The repulsive term $V_1$ introduced in BW91$^*$ raises the pocket minimum relative to BW91, bringing the capture-pocket depth into close agreement with the SkM$^*$ energy-density functional ($B_{\rm cap}=7.15$ MeV from SkM$^*$ and 7.11 MeV from BW91$^*$). Taken together, these results validate BW91$^*$ as a reliable input for subsequent capture and fusion cross-section calculations. 

 \subsection{ Asymmetric Gaussian Distribution }

In this work, the most probable barrier height $B_m$ is expressed as,
\begin{equation}
	B_m=B_0 - 0.1 Z^{1/3} (1+\eta Z^{1/3}), 
		\label{eq:Bm}
\end{equation}
where $Z=Z_1+Z_2$ is the total charge of the compound nucleus and $\eta=|A_1-A_2|/(A_1+A_2)$ characterizes the mass asymmetry of the entrance channel. The second term in Eq.~\eqref{eq:Bm} accounts for the fact that the barrier height $B_0$ in the BW91$^*$ potential is slightly higher than the original one and that the most probable barrier height is lower than the average barrier height due to the asymmetric distribution (which will be discussed in the next subsection).

\begin{figure}[htbp]
	\setlength{\abovecaptionskip}{ 0.2cm}
	\includegraphics[angle=0,width=0.6\textwidth]{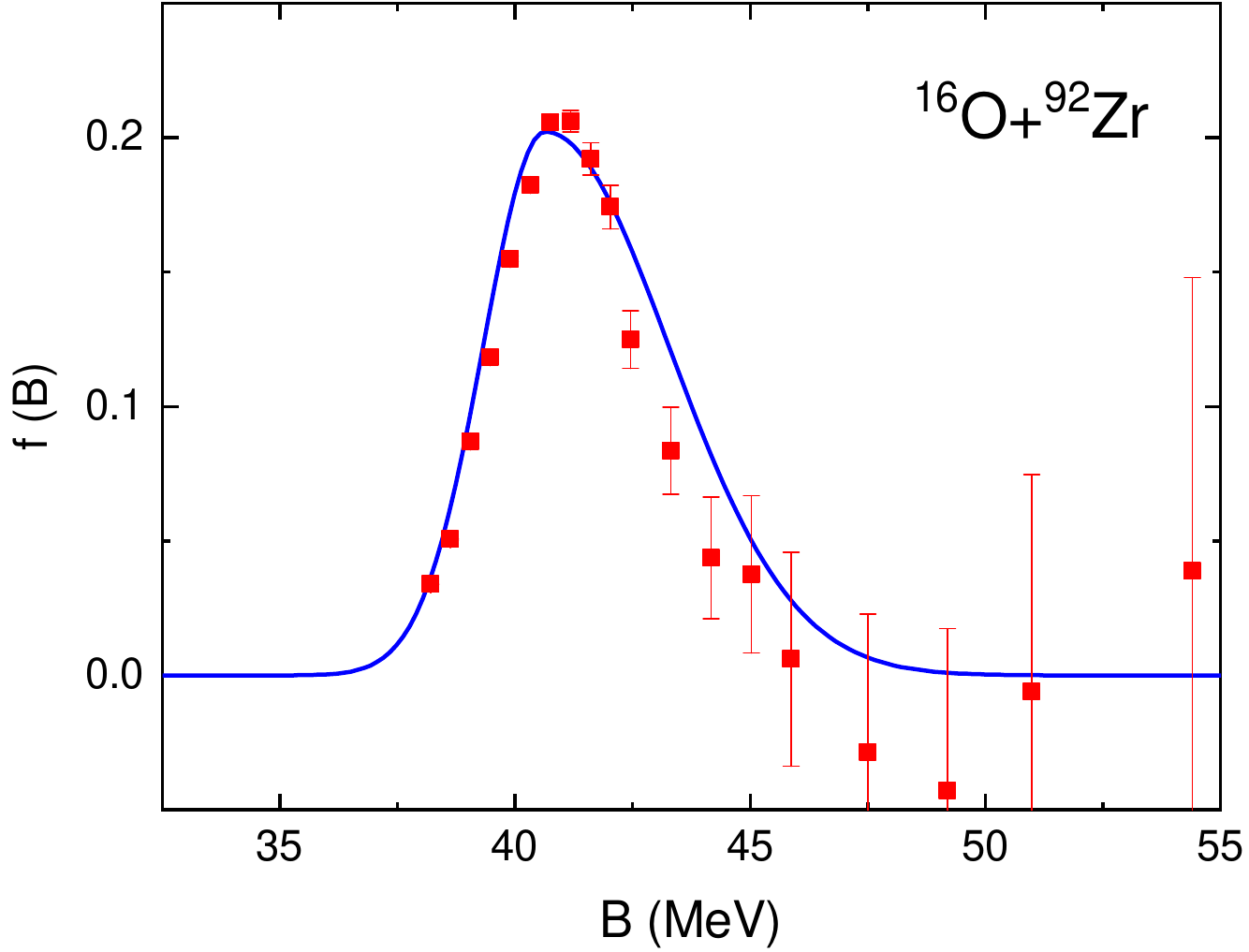}	
	\caption{Barrier distribution $f(B)$ for the reaction $^{16}$O + $^{92}$Zr. The red squares with error bars represent the barrier distribution extracted from the experimental capture excitation function\cite{Newton2001Experimental}. The blue solid curve denotes the ECC2 prediction. }
	\label{fig:bdis}
\end{figure}

Considering that the sub-barrier capture cross sections are strongly influenced by nuclear structure effects such as nuclear deformations, we introduce a deformation-dependent distribution width in this work. The left and right widths in the barrier distribution function $f(B)$ [see Eq.(5)] are given by \( \Delta_1 = \tfrac{3}{2} W\) and \( \Delta_2 = 2 \Delta_1\), where $W$ denotes the standard deviation of the Gaussian function in empirical barrier distribution (EBD2) method \cite{Wang_2025,Wang_2026},
\begin{equation}
	W=c_0 (1+w_{d}) + c_1 B_m \sqrt{w_1^2+w_2^2+w_0^2},
\end{equation} 
where $w_{d}=\sum_i |\beta_{2i}| A_i^{1/3}$ and $w_i = A_i^{1/3} \beta_{2i}^2/(4\pi)$, with the mass numbers $A_1$ and $A_2$ of the reaction partners, and their quadrupole deformation parameters $\beta_2$ taken from the WS4 model \cite{WS4} for prolate nuclei heavier than $^{16}$O. The term $w_0=(B_m+Q)/c_2$ is introduced to account for the dynamical effects associated with the excitation energy at the capture position \cite{Yao24}, and $Q$ denotes the reaction $Q$-value in fusion. The parameter values $c_0=0.63$ MeV, $c_1=0.015$, and $c_2=33.0$ MeV are identical to those adopted in EBD2 \cite{Wang_2025,Wang_2026}.

Figure~\ref{fig:bdis} shows the fusion barrier distribution for $^{16}$O + $^{92}$Zr. The curve denotes the result of ECC2 and the squares denote the extracted distribution from the measured capture excitation function \cite{Newton2001Experimental}. One sees that both the peak and the left width of the extracted barrier distribution are reproduced remarkably well by ECC2.

 \subsection{ Effective Barrier Curvature }

\begin{figure}[htbp]
	\setlength{\abovecaptionskip}{0.0cm}
	\includegraphics[angle=0,width=0.65\textwidth]{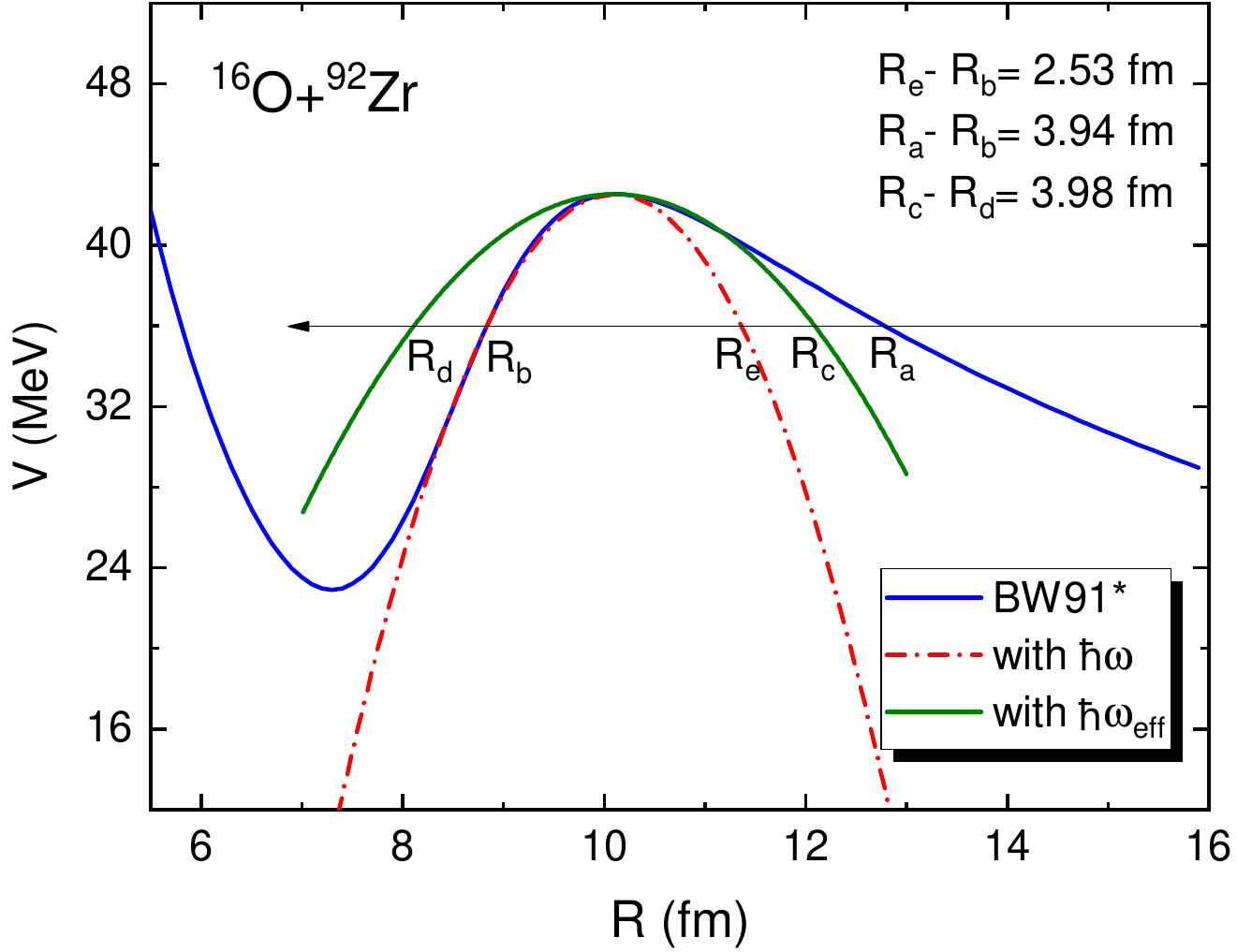}

	\caption{Nucleus-nucleus potential $V(R)$ for the reaction $^{16}$O + $^{92}$Zr. The blue solid line denotes the BW91$^*$ potential. The red dot-dashed and green solid lines denote the parabolic approximations with $\hbar\omega$ and $\hbar\omega_{\rm eff}$, respectively. The horizontal black line marks the reference energy $E_{\rm c.m.}=0.85\,B_0$. The outer turning points ($R_a$, $R_c$, $R_e$) and inner turning points ($R_b$, $R_d$) of the different curves at this energy are labeled, and the corresponding barrier widths $R_a-R_b$, $R_e-R_b$, and $R_c-R_d$ are indicated in the top-right corner.}

\label{fig:barrierwidth}
\end{figure}

In the traditional ECC calculations, the barrier curvature is usually evaluated with
\begin{equation}
	\hbar \omega =  \sqrt{-\frac{\hbar^2}{\mu} \frac{\partial^2 V}{\partial R^2}} \bigg|_{R=R_0} .
	\label{eq:hw}
\end{equation}
Here, $R_0$ denotes the position of the barrier maximum. For realistic potentials, however, the width of the Coulomb barrier is significantly larger than that predicted by Eq.~\eqref{eq:hw}, mainly because of the long tail of the Coulomb potential.  In Fig. \ref{fig:barrierwidth}, we show the BW91$^*$ potential for $^{16}$O + $^{92}$Zr. The dashed curve denotes the corresponding parabolic potential with the standard curvature according to Eq.~\eqref{eq:hw}. One sees that the parabolic potential with the standard curvature is significantly narrower than the BW91$^*$ potential in the sub-barrier region due to the long-range Coulomb potential. For example, at $E_{\rm c.m.}=0.85 B_0$, the barrier width from the standard $\hbar\omega$ is $R_e-R_b=2.53$ fm, which is markedly smaller than the value $R_a-R_b=3.94$ fm obtained from the ${\rm BW91^*}$ potential. Such an artificially narrow barrier tends to enhance the Hill-Wheeler penetration probability at sub-barrier energies.  

 To better reproduce the width of the Coulomb barrier, we therefore adopt an effective barrier curvature $\hbar \omega_{\rm eff}$ defined as
\begin{equation}
	\hbar \omega_{\rm eff} =  \sqrt{-\frac{1}{2} \frac{\partial^2 V}{\partial R^2}} \bigg|_{R=R_0} .
	\label{eq:hweff}
\end{equation} 

The coefficient $1/2$ ${\rm MeV\cdot fm^{2}}$ in Eq.~\eqref{eq:hweff} is fixed for all reactions in the calculations. The green solid curve in Fig.~\ref{fig:barrierwidth} denotes the parabolic potential based on $\hbar \omega_{\rm eff}$, and its barrier width $R_c-R_d=3.98$ fm is in good agreement with that of the ${\rm BW91^*}$ potential ($R_a-R_b$). The turning points $R_a$, $R_b$, $R_c$, $R_d$, and $R_e$ are labeled in Fig.~\ref{fig:barrierwidth}. With the effective barrier curvature, the over-prediction of the penetration probability at sub-barrier energies in ECC can be improved.

\section{Results and Discussion}

  \begin{figure}[htbp]
	\setlength{\abovecaptionskip}{ 0.0cm}
	\includegraphics[angle=0,width=0.7\textwidth]{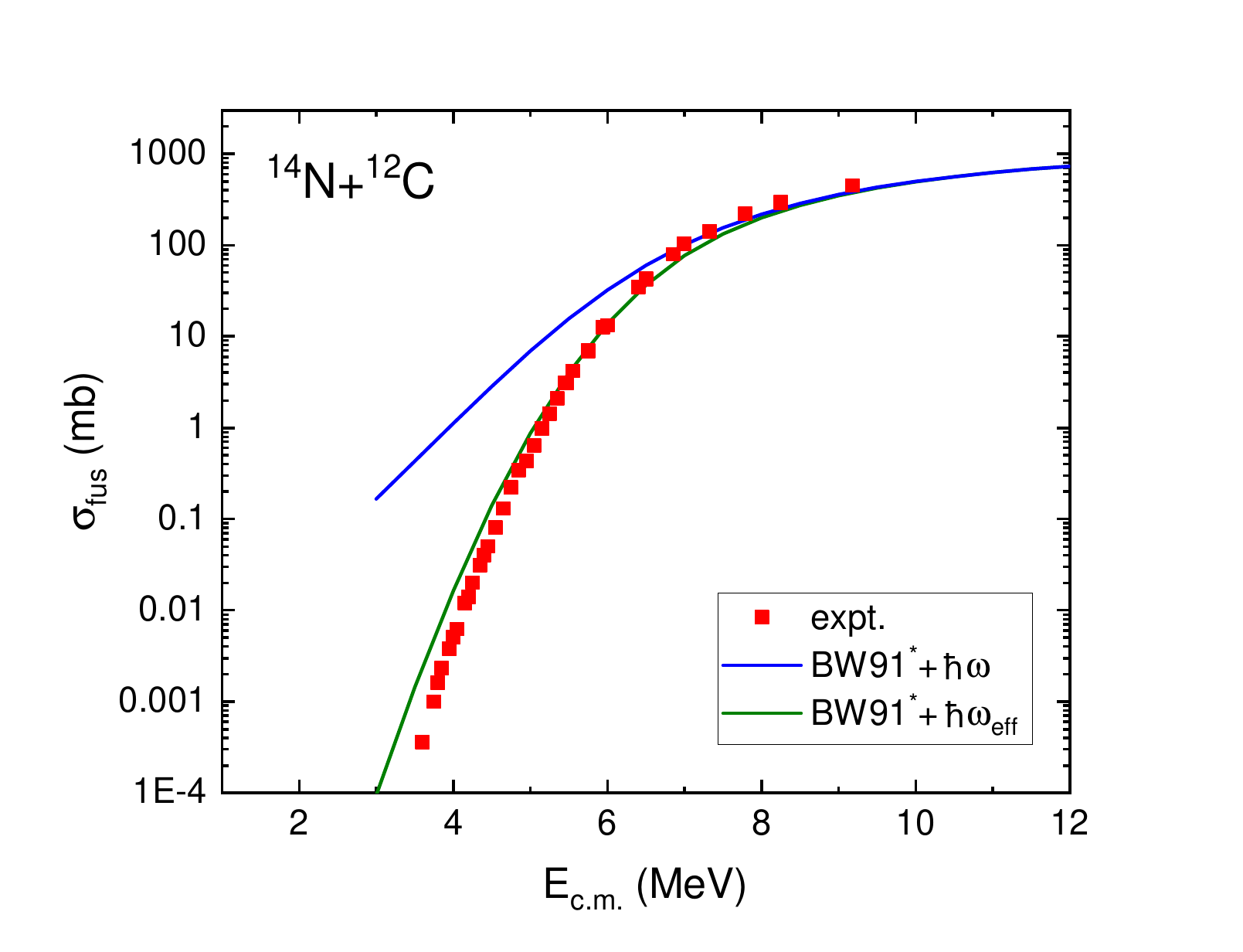}
	\caption{Fusion excitation function for reaction $^{14}$N + $^{12}$C. The blue and green curves denote the results with the standard curvature and the effective curvature, respectively. The red solid squares denote the experimental data taken from \cite{Switkowski1977Fusion}.}
	\label{fig:curvature}
\end{figure}

Before presenting the systematic results, we first illustrate the impact of the effective barrier curvature introduced in Sec.~II.D. As shown in Fig.~\ref{fig:curvature}, the fusion (i.e. capture) cross sections for the reaction $^{14}$N + $^{12}$C exhibit a clear sensitivity to the choice of curvature. The blue curve, obtained with the standard curvature $\hbar\omega$, overestimates the cross sections at sub-barrier energies. This overestimation can be attributed to the artificially narrow parabolic barrier shown in Fig.~\ref{fig:barrierwidth}, which enhances the Hill--Wheeler penetration probability below the barrier. When the effective curvature $\hbar\omega_{\rm eff}$ is adopted instead (green curve), the broader barrier width suppresses the sub-barrier penetration, and the calculated cross sections are brought into substantially better agreement with the experimental data~\cite{Switkowski1977Fusion}. These results demonstrate that the effective barrier curvature corrects the systematic overestimation inherent in the standard Hill--Wheeler formula and validate its use throughout the subsequent ECC2 calculations.

\begin{figure}[htbp]
    \setlength{\abovecaptionskip}{ 0.0cm}
    \includegraphics[angle=0,width=1\textwidth]{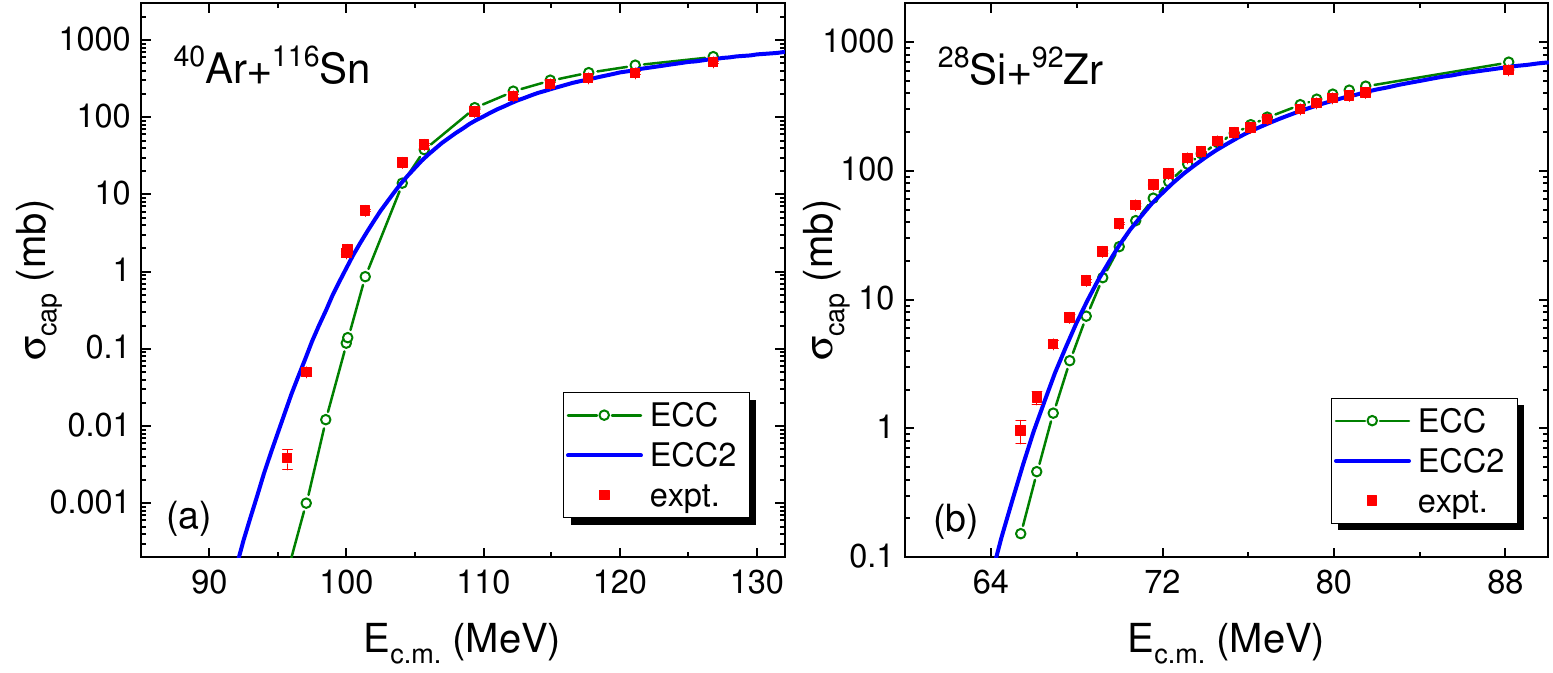}
    \caption{Capture cross sections $\sigma_{\rm cap}$ as a function of the center-of-mass energy $E_{\rm c.m.}$ for the reactions (a) $^{40}$Ar + $^{116}$Sn and (b) $^{28}$Si + $^{92}$Zr. The red solid squares denote the experimental data taken from Refs. \cite{Reisdorf1985Fusion,Newton2001Experimental}. The green open circles connected by lines denote the ECC model predictions, and the blue solid curves denote the ECC2 results.}
  \label{fig:6}
 \end{figure}

We now compare the full ECC2 model with the earlier ECC version~\cite{Wang_2017}. It should be noted that the two models differ not only in the barrier curvature but also in the bare interaction potential (BW91$^*$) and the barrier-distribution widths, so the net change in the predicted cross sections arises from the combined effect of all three improvements. As representative examples, Fig.~\ref{fig:6} presents the capture excitation functions for $^{40}$Ar + $^{116}$Sn and $^{28}$Si + $^{92}$Zr, in which the ECC2 predictions are compared with the ECC results and the experimental data. In both systems, the ECC2 results (blue solid curves) are in good overall agreement with the experimental data~\cite{Reisdorf1985Fusion,Newton2001Experimental} across the measured energy range. In contrast, the ECC model (green open circles connected by lines) systematically underestimates the cross sections at sub-barrier energies in both systems.

\begin{figure}[htbp]
	\setlength{\abovecaptionskip}{ 0.0cm}
	\includegraphics[angle=0,width=1\textwidth]{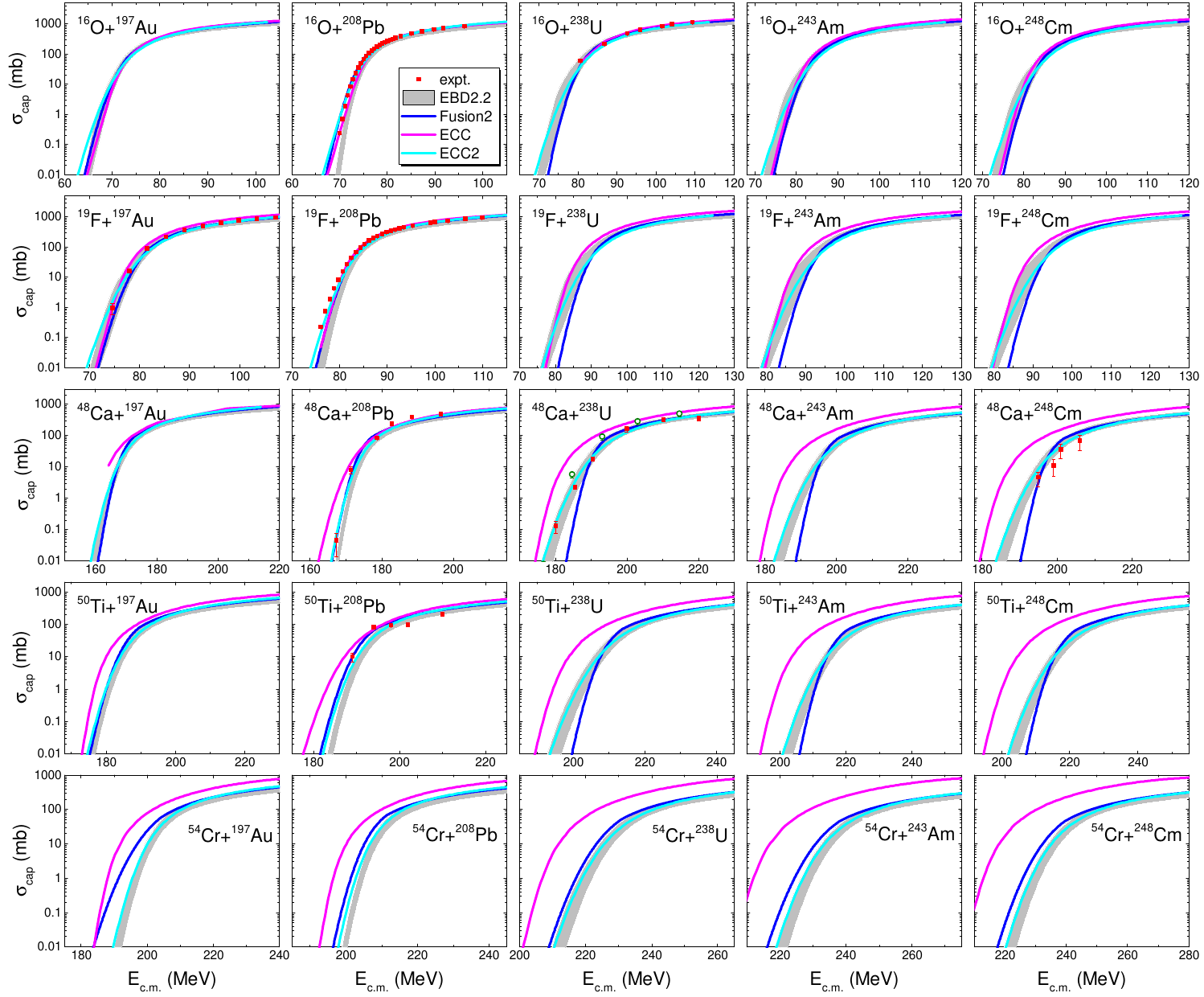}
	\caption{Predicted capture excitation functions for various projectile--target combinations, arranged as a $5\times 5$ matrix ranging from $^{16}$O- to $^{54}$Cr-induced reactions on the targets $^{197}$Au, $^{208}$Pb, $^{238}$U, $^{243}$Am, and $^{248}$Cm. The red squares and green circles represent the experimental data from Refs.~\cite{Morton1999Coupled,Hinde1999Limiting,Prokhorova_2008,Clerc_1984,Nishio2012Fusion,Itkis2022Experimental}. The shaded grey band and the solid magenta, blue, and cyan curves denote the calculated results of the EBD2.2, Fusion-v2, ECC (2017), and ECC2 models, respectively.}
	\label{fig:25matrix}
\end{figure}

The complete set of 25 excitation functions covering a wide range of projectile--target combinations, from medium systems induced by $^{16}$O and $^{19}$F to heavy systems induced by $^{48}$Ca, $^{50}$Ti, and $^{54}$Cr, is displayed in Fig.~\ref{fig:25matrix}. The static deformations of the reaction partners in these systems are also quite different, including oblate, spherical and prolate shapes. The experimental data are taken from Refs.~\cite{Morton1999Coupled,Hinde1999Limiting,Prokhorova_2008,Clerc_1984,Nishio2012Fusion,Itkis2022Experimental}. The ECC2 predictions are compared with those of EBD2.2, the universal Wong formula (Fusion-v2), and ECC (2017). Overall, ECC2 reproduces the experimental excitation functions across the entire mass range with a single set of globally fixed parameters. Notably, for hot-fusion reactions such as $^{48}$Ca + $^{238}$U and $^{48}$Ca + $^{248}$Cm, ECC2 shows markedly better agreement with the data than ECC (2017) and Fusion-v2, particularly at sub-barrier and near-barrier energies, owing to the improved barrier curvature, the refined potential and the deformation-dependent distribution width.  For superheavy systems with shallow capture pockets, not all partial waves that overcome the Coulomb barrier contribute to capture, as discussed by Zagrebaev {\it et al.}~\cite{Zag01} in terms of a critical angular momentum $L_{\rm cr}$. Following the universal Wong formula~\cite{Wang2025Universal}, this effect is accounted for in ECC2 by a suppression factor $F_{\rm DIS}$, which was also adopted in the ECC calculations~\cite{Zhang_2025} and was found to significantly improve the above-barrier cross section description.

\begin{eqnarray}
	F_{\rm DIS}= \frac{1}{2}\left [ 1+{\rm erf}(\sqrt{B_{\rm cap}/c_d}-1) \right ],
\end{eqnarray}   
with $c_d=2.0$ MeV. The average barrier radius is written as $R_B=R_0 F_{\rm DIS}$ with $R_0$ being the barrier radius in the BW91$^*$ potential. For light and intermediate fusion systems, $R_B \approx R_0$ due to the deep capture pocket. For super-heavy systems with shallow pockets, the values of $R_B$ are significantly smaller than those of $R_0$, which evidently reduces the capture cross sections at above-barrier energies.

    \begin{figure}[htbp]
	\setlength{\abovecaptionskip}{0.0cm}
	\includegraphics[angle=0,width=0.75\textwidth]{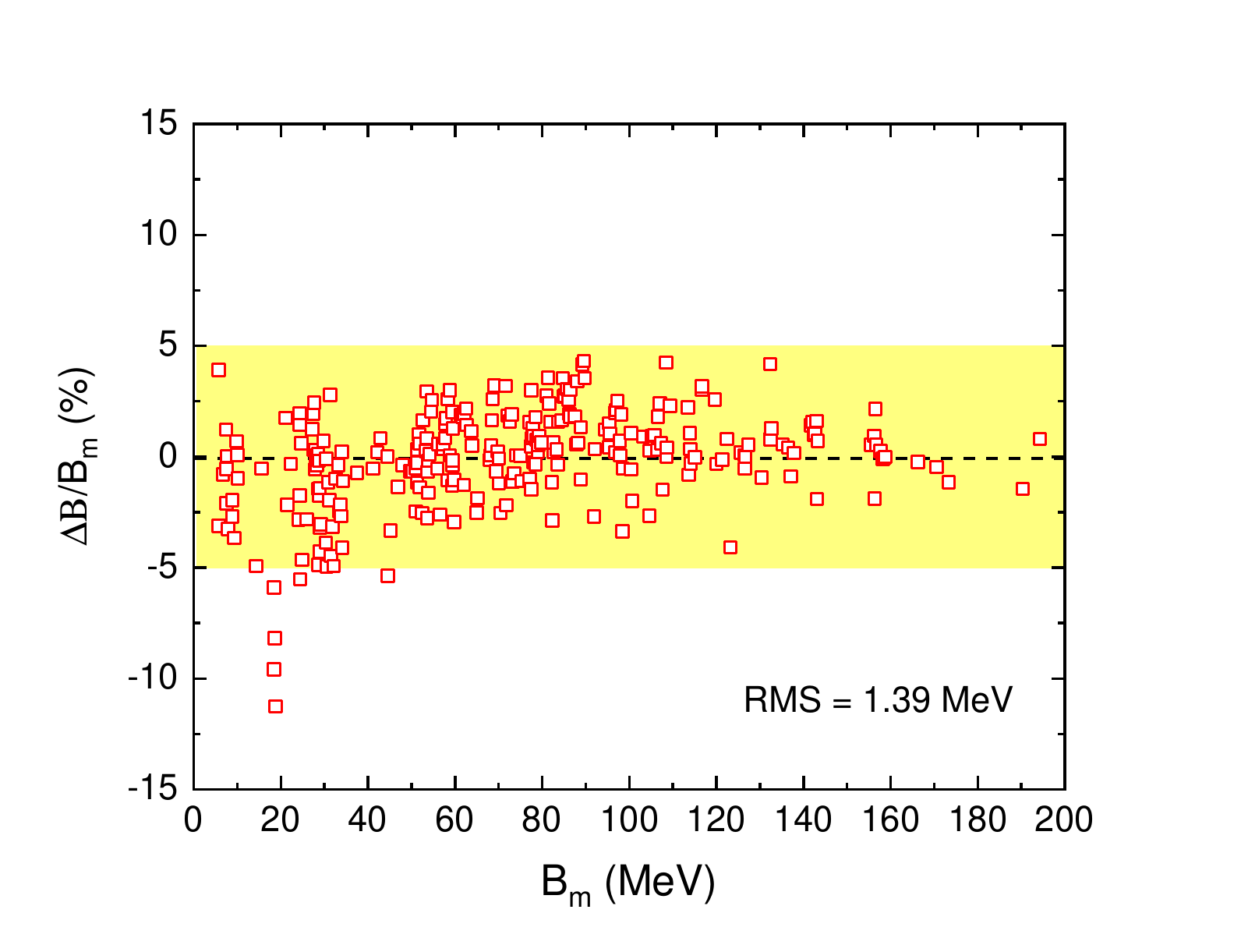}
	\caption{Relative deviation between the most probable barrier height $B_{m}$ in ECC2 and the extracted barrier heights. The yellow band highlights the $\pm 5\%$ range, and the RMS deviation is 1.39 MeV.}
	\label{fig:dVBvsBm}
\end{figure}

Table~\ref{tab:barrierparams} lists the barrier parameters corresponding to the excitation functions shown in Fig.~\ref{fig:25matrix}. The ECC2 most probable barrier height $B_m$ agrees with the EBD2.2 barrier height $V_B$ within a relative deviation of 0.6\% for all systems, and the ECC2 capture-pocket depth $B_{\rm cap}$ is in close agreement with that from the universal Wong formula with the SkM$^*$ energy-density functional. These consistencies provide quantitative support for the agreement seen in the excitation functions.

\begin{table*}[htbp]
\centering
\caption{Capture barrier height $V_B$ from EBD2.2, together with the most probable barrier height $B_m$ predicted by ECC2, for the 25 fusion reactions considered in this work. Here, the capture-pocket depths $B_{\rm cap}$ from ECC2 and those from the universal Wong formula based on the SkM$^{*}$ energy-density functional \cite{Bartel1982} are also listed for comparison. All values are given in MeV.}
\label{tab:barrierparams}
\renewcommand{\arraystretch}{1.15}
\setlength{\tabcolsep}{4pt}
\begin{tabular}{l c c c c @{\hspace{1.2em}} l c c c c}
\hline\hline
Reaction & $V_B$ & $B_m$ & $B_{\rm cap}^{\rm ECC2}$ & $B_{\rm cap}^{\rm {SkM}^{*}}$ & Reaction & $V_B$ & $B_m$ & $B_{\rm cap}^{\rm ECC2}$ & $B_{\rm cap}^{\rm {SkM}^{*}}$ \\
\hline
$^{16}$O + $^{197}$Au  & 72.00  & 72.20  & 16.53 & 16.65 & $^{48}$Ca + $^{197}$Au & 168.89 & 168.67 & 7.45 & 7.51 \\
$^{16}$O + $^{208}$Pb  & 74.35  & 74.18  & 16.49 & 16.78 & $^{48}$Ca + $^{208}$Pb & 174.16 & 173.48 & 7.11 & 7.15 \\
$^{16}$O + $^{238}$U   & 80.85  & 81.25  & 15.90 & 16.35 & $^{48}$Ca + $^{238}$U  & 191.03 & 190.46 & 5.92 & 5.93 \\
$^{16}$O + $^{243}$Am  & 83.30  & 83.68  & 15.50 & 15.80 & $^{48}$Ca + $^{243}$Am & 196.85 & 196.15 & 5.48 & 5.50 \\
$^{16}$O + $^{248}$Cm  & 83.79  & 84.20  & 15.54 & 15.95 & $^{48}$Ca + $^{248}$Cm & 198.22 & 197.49 & 5.44 & 5.42 \\
$^{19}$F + $^{197}$Au  & 80.20  & 80.44  & 15.70 & 15.61 & $^{50}$Ti + $^{197}$Au & 185.90 & 185.52 & 6.20 & 6.28 \\
$^{19}$F + $^{208}$Pb  & 82.78  & 82.67  & 15.59 & 15.62 & $^{50}$Ti + $^{208}$Pb & 191.69 & 190.81 & 5.91 & 6.01 \\
$^{19}$F + $^{238}$U   & 89.81  & 90.58  & 14.89 & 15.04 & $^{50}$Ti + $^{238}$U  & 210.24 & 209.48 & 4.76 & 4.87 \\
$^{19}$F + $^{243}$Am  & 92.61  & 93.28  & 14.48 & 14.52 & $^{50}$Ti + $^{243}$Am & 216.77 & 215.75 & 4.31 & 4.38 \\
$^{19}$F + $^{248}$Cm  & 93.11  & 93.87  & 14.52 & 14.63 & $^{50}$Ti + $^{248}$Cm & 218.23 & 217.22 & 4.27 & 4.36 \\
$^{54}$Cr + $^{197}$Au & 202.09 & 201.36 & 5.08  & 5.16  & $^{54}$Cr + $^{243}$Am & 235.84 & 234.25 & 3.25 & 3.29 \\
$^{54}$Cr + $^{208}$Pb & 208.39 & 207.11 & 4.81  & 4.90  & $^{54}$Cr + $^{248}$Cm & 237.43 & 235.82 & 3.27 & 3.27 \\
$^{54}$Cr + $^{238}$U  & 228.69 & 227.41 & 3.68  & 3.80  &                         &        &         &       &       \\
\hline\hline
\end{tabular}
\end{table*}

 As the most probable barrier height $B_m$ is the central parameter of barrier distribution, we next assess its reliability across the full range of systems. To this end, the ECC2 barrier height $B_m$ from the BW91$^*$ potential via Eq.~\eqref{eq:Bm}, is compared with the extracted barrier heights from the measured excitation functions for 367 reaction systems~\cite{Chen23AtData}. As can be seen from Fig.~\ref{fig:dVBvsBm}, the relative deviation from the experimental values is centered near zero and falls within $\pm 5\%$ for nearly all systems, with an rms deviation of 1.39 MeV.  No systematic dependence on $Z_1Z_2$ is observed. Taken together, these results demonstrate that the globally fixed barrier-height prescription of ECC2 provides an accurate and unbiased description over the entire mass range considered.

\section{Summary}

In summary, we have developed a universal empirical coupled channel model (ECC2) for the systematic calculation of capture cross sections in heavy-ion fusion reactions. The model was constructed to address two limitations of earlier ECC calculations: the model dependence of the bare barrier and the overly narrow parabolic barrier used in the standard Hill--Wheeler penetration probability.  ECC2 treats these issues through (i) the modified Broglia--Winther (BW91$^*$) potential with a newly introduced repulsive term $V_1$, (ii) an effective barrier curvature $\hbar\omega_{\rm eff}$, and (iii) an asymmetric Gaussian barrier distribution with EBD2-based deformation-dependent width parameters. All parameters are globally fixed.

The effective barrier curvature was shown to correct the systematic overestimation of sub-barrier cross sections inherent in the standard Hill--Wheeler formula. The full ECC2 model was then applied to 25 fusion reactions covering a broad range of projectile--target combinations. The predicted capture excitation functions are consistent with the EBD2.2 predictions for all systems and agree well with the available experimental data. For hot-fusion reactions such as $^{48}$Ca + $^{238}$U and $^{48}$Ca + $^{248}$Cm, ECC2 shows better agreement with the data than ECC (2017) and Fusion-v2, owing to the refined potential and the effective barrier curvature, as well as the suppression factor $F_{\rm DIS}$. The most probable barrier height $B_m$ agrees with experimental values for 367 systems to within an rms deviation of 1.39 MeV, and the corresponding barrier parameters are in close agreement with those from the EBD2.2 model and the universal Wong formula with the SkM$^*$ energy-density functional, respectively.

The present ECC2 model provides a reliable and computationally efficient tool for predicting capture cross sections in the synthesis of superheavy nuclei, which constitutes the essential first step for evaporation residue cross section calculations within the DNS framework. Further applications to neutron-rich projectile--target combinations and to yet-unmeasured Ti- and Cr-induced reactions will be useful for assessing the predictive power of the model in future superheavy-element experiments.

\section*{Acknowledgments}
This work was supported by Guangxi "Bagui Scholar" Teams for Innovation and Research Project, Guangxi Natural Science Foundation (No. 2026GXNSFBA00640195), National Natural Science Foundation of China (Nos. 12265006, U1867212, 12475121, 12435008, 12375127), Strategic Priority Research Program of the Chinese Academy of Sciences (Grant No. XDB1550100) and the National Key R\&D Program of China (Contract No. 2023YFA1606503). Helpful discussions with members of the Superheavy Nuclei Theory Collaboration (SNTC) are highly appreciated. Online calculations with ECC2 are available on http://www.imqmd.com/fusion/ECC2/

\bibliographystyle{apsrev4-2}
\bibliography{ECC}

\end{document}